\documentclass[12pt]{article}

\usepackage{times}
\usepackage{graphicx}
\usepackage{setspace}
\usepackage{color}
\usepackage{amsfonts,amsmath,amsthm}

\newcommand{\bket}[1]{\left<#1\right>}

\newcommand{\ket}[1]{\left|#1\right>}

\newenvironment{sciabstract}{%
\begin{quote} \bf}
{\end{quote}}

\newcounter{lastnote}
\newenvironment{scilastnote}{%
\setcounter{lastnote}{\value{enumiv}}%
\addtocounter{lastnote}{+1}%
\begin{list}%
{\arabic{lastnote}.}
{\setlength{\leftmargin}{.22in}}
{\setlength{\labelsep}{.5em}}}
{\end{list}}

\title{Quantum back-action of variable-strength measurement} 

\author
{M. Hatridge$^{1\ast}$ and S. Shankar,$^{1}$ M. Mirrahimi,$^{1,2}$ F. Schackert,$^{1}$\\
K. Geerlings,$^{1}$ T. Brecht,$^{1}$ K.~M. Sliwa,$^{1}$ B. Abdo,$^{1}$\\
L. Frunzio,$^1$ S.~M. Girvin, $^1$ R.~J. Schoelkopf, $^1$ M.~H. Devoret$^{1}$\\
\\
\normalsize{$^{1}$ Department of Applied Physics and Physics, Yale University, New Haven, Connecticut 06520, USA,}\\
\normalsize{$^{2}$ INRIA Paris-Rocquencourt, Domaine de Voluceau, B.P. 105, 78153 Le Chesnay Cedex, France}\\
\normalsize{$^\ast$To whom correspondence should be addressed; E-mail:  michael.hatridge@yale.edu.}
}

\date{}

\begin{document} 




\maketitle 


\begin{sciabstract}
Measuring a quantum system can randomly perturb its state. The strength and nature of this ‘back-action’ depends on the quantity which is measured. In a partial measurement performed by an ideal apparatus, quantum physics predicts that the system remains in a pure state whose evolution can be tracked perfectly from the measurement record. We demonstrate this property using a superconducting qubit dispersively coupled to a cavity traversed by a microwave signal.  The back-action on the qubit state of a single measurement of both signal quadratures is observed and shown to produce a stochastic operation whose action is determined by the measurement result.  This accurate monitoring of a qubit state is an essential prerequisite for measurement-based feedback control of quantum systems.
\end{sciabstract}

While the behavior (`state collapse') of a quantum system subject to an infinitely-strong, i.e. projective, quantum non-demolition (QND) measurement
is textbook physics, the subtlety and utility of finite-strength, i.e. partial, measurement phenomena
are neither widely appreciated nor commonly verified experimentally.  Standard quantum measurement theory puts forward the principle that observing a system induces a decoherent evolution proportional to the measurement strength~\cite{Brune1996, Haroche2006, Schuster2005, Gambetta2008, Sears2012}.  Thus, partial measurement is often associated with partial decoherence of the state of a quantum system.  However,  this measurement-induced degradation occurs only if the measurement is inefficient ``informationally'', i.e. if only a portion of the measurement's information content is available to the observer for use in reconstructing the new state of the system.  

 If, instead, the measurement apparatus is entirely efficient, the new state of the quantum system can be perfectly reconstructed.  This  outcome-dependent revision of the system's imposed initial conditions constitutes a fundamental quantum effect called ``measurement back-action''~\cite{Carmichael1993,Haroche2006,Wiseman2009,Korotkov2011}. Although the system's evolution under measurement is erratic, hence the measurement outcome cannot be predicted in advance,  the measurement record faithfully reports the perturbation of the system after the fact.

We utilize the powerful, combined qubit-cavity architecture, Circuit Quantum Electrodynamics (cQED)~\cite{Blais2004,Wallraff2004}, which allows for rapid, repeated Quantum Non-Demolition (QND)~\cite{Braginsky1980,Grangier1998} superconducting qubit measurement~\cite{Palacios-Laloy2010, Schoelkopf2008, Hoffman2011, Mariantoni2011, Riste2012, Vijay2012 }.   The cavity output is monitored in real time using a phase-preserving amplifier working near the quantum-limit, where the noise is only caused by the fundamental quantum fluctuations of the electrodynamic vacuum~\cite{Caves1982}. The decision to read out our qubit using coherent states of the resonator has two important consequences. First, the outcomes of a partial measurement form a quasi-continuum, unlike the set of discrete answers obtained from a projective measurement.  Second, measuring both quadratures of the signal leads to two-dimensional diffusion of the direction of the qubit effective spin.  We show that the choice of measurement apparatus and of measurement strength both affect the evolution of a quantum system, but neither results in degradation of the system's state if the measurement is informationally efficient.  Such precise knowledge of the measurement back-action is a necessary prerequisite for general feedback control of quantum systems.

Our superconducting qubit is a transmon~\cite{Schreier2008}, consisting of two Josephson junctions in a closed loop, shunted by a capacitor to form an anharmonic oscillator. The two lowest energy states,  $ (|g\rangle$ and $|e\rangle )$, are the logical states of the qubit. The qubit is dispersively coupled to a compact resonator, which is further asymmetrically coupled to input and output transmission lines (Fig. 1B, C),  determining the resonator bandwidth ($\kappa/2\pi=5.8$~MHz).  To measure a qubit prepared in initial state $\ket{\psi}=c_g\ket{g}+c_e\ket{e}$, a microwave pulse of duration $T_m \gg 1/\kappa$ is applied to the resonator.  The state dependent shift of the resonator frequency ($\chi/2\pi=5.4$~MHz) results in an entangled state of the qubit and pulse $\ket{\Psi}=c_g\ket{g}\otimes\ket{\alpha_g}+c_e\ket{e}\otimes\ket{\alpha_e}$, where $\ket{\alpha_{g,e}}$ refer to the coherent state after traversing the resonator.

Amplification is required to convert the pointer state $\ket{\Psi}$ into a macroscopic signal that can be processed and recorded with standard instrumentation. In our case, the pulse having traversed the resonator is amplified using a linear, phase-preserving amplifier with gain G, which can be seen as multiplying the average photon number in $\ket{\alpha_{g,e}}$(see Fig.~1B). For dynamical range considerations, our Josephson amplifier is operated in this experiment with a gain $G=12.5$~dB and bandwidth of 6 MHz, adding close to the minimum amount of noise allowed by quantum mechanics~\cite{Bergeal2010, Bergeal2010a,Abdo2011,Roch2012}.  The added quantum fluctuations are due to a second, ``idler'', input~\cite{Caves1982}.  A measurement of both quadratures of the output mode results in an outcome, denoted $(I_m,Q_m)$, which is then used to determine the new state of the qubit after measurement (see Fig.~1A).  As has been shown in~\cite{Korotkov2011}, and detailed in the supplementary material, this outcome contains all information necessary to perfectly reconstruct the new state of the qubit.  Remarkably, the additional quantum fluctuations introduced during amplification enter in the measurement back-action on the qubit without impairing our knowledge of it.

We first demonstrate projective qubit readout by strongly measuring the qubit using an 8~$\mu$s pulse with the drive power set so that the average number of photons in the resonator during the pulse was $\bar n=5$ (Fig 2).  Selected individual measurement records for the qubit are shown in Fig. 2B. The data are digitized with a sampling time of $20$~ns and smoothed with a binomial filter with $T_m=240$~ns width, corresponding to $8$ cavity lifetimes, and scaled by the experimentally determined standard deviation ($\sigma$).  The highlighted trace shows clear quantum jumps in the qubit state, which are identified by vertical black dotted lines indicating $4\sigma$ deviations from the current qubit state. The $8$\% equilibrium qubit excited state population is consistent with other measurements of superconducting qubits~\cite{Corcoles2011}.  By counting the number of up and down transitions in $25,000$ traces with no qubit excitation pulse, we calculate $T_1 \le 3.1~\mu\text{s}$.  Although we fail to resolve pairs of transitions separated by much less than our filter time constant, this method for estimating $T_1$ yields a value in good agreement with the value $T_1=2.8~\mu\text{s}$ calculated from fitting an exponential to the averaged trajectory of all traces.  Further, the average qubit polarization did not vary over $8$~$\mu$s of continuous measurement, nor did $T_1$ diminish with larger readout amplitude up to $\bar n \simeq 15$, demonstrating the QND nature of our readout.

 Histograms of the scaled $I_m$ component of the outcome for the first $240$~ns of measurement after a qubit rotation by $\theta=0,\pi/2,\pi$ are shown in Fig. 2C. The ground and excited distributions are separated by 4.8 standard deviations, corresponding to a measurement fidelity of $98$\% when $I_m=0$ is used as the discrimination threshold.  We emphasize that the discreteness of the $z$ measurement of the transmon circuit, illustrated by the bimodality of the histogram, is here due only to the quantum nature of the circuit and not to any nonlinearity of the readout.  Thus, this measurement of a continuous, unbounded pointer state is exactly equivalent to the Stern-Gerlach experiment. These strong, high-fidelity measurements allow us to perform precise tomography, and to prepare the qubit in a known state by measurement.  We next use these tools to quantify measurement back-action of partial measurement on the qubit state. 

The qubit evolution due to partial measurement can be precisely calculated from the complete measurement record using the quantum trajectory approach~\cite{Carmichael1993, Wiseman2009}, but this is computationally intensive.  Instead, we calculate the back-action from the average output over the time $T_m$, as in~\cite{Korotkov2011}.  Provided that the measurement time is short compared to the qubit coherence times $T_1$ and $T_2$, and long compared to the cavity lifetime and amplifier response time, this approach allows the qubit to be tracked without degradation.  In this experiment, $T_m=240$~ns, which is shorter than $T_1=2.8~\mu$s and $T_{2R}=0.7-2.0~\mu$s, and much longer than the cavity lifetime and JPC response time of 30~ns.  

Assuming the qubit is initially polarized along +y-axis, we calculate the final qubit Bloch vector $(x_f, y_f, z_f)$ as a function of measurement outcome $(I_m,Q_m)$ (see detailed derivation in the Supplementary Materials) to be :
\begin{align}\label{eq:final_bloch_vector_main}
x_f^\eta(I_m,Q_m)&=\text{sech}\left(\frac{I_m\bar I_m}{\sigma^2}\right)\sin{\left(\frac{Q_m\bar I_m}{\sigma^2}+\frac{\bar Q_m\bar I_m}{\sigma^2}\left(\frac{1-\eta}{\eta}\right)\right)}e^{-\frac{\bar I_m^2}{\sigma^2}\left(\frac{1-\eta}{\eta}\right)},\notag\\
y_f^\eta(I_m,Q_m)&=\text{sech}\left(\frac{I_m\bar I_m}{\sigma^2}\right)\cos{\left(\frac{Q_m\bar I_m}{\sigma^2}+\frac{\bar Q_m\bar I_m}{\sigma^2}\left(\frac{1-\eta}{\eta}\right)\right)}e^{-\frac{\bar I_m^2}{\sigma^2}\left(\frac{1-\eta}{\eta}\right)},\\
z_f^\eta(I_m)&=\tanh{\left(\frac{I_m\bar I_m}{\sigma^2}\right)},\notag
\end{align}

where $\bar I_m$ and $\bar Q_m$ and $\sigma$ define the center and standard deviation of the outcome distributions, and $\eta$ is the quantum efficiency of the amplification chain (Fig.~1A).  In this theory, we neglect the effect of qubit decoherence and losses before amplification.  In the limit of a perfectly efficient amplification $(\eta=1)$, we see that the length of the Bloch vector is unity, irrespective of outcome.  The parameter $\bar I_m/\sigma$ can be identified as  the apparent measurement strength since the measurement becomes more strongly projective as $\bar I_m/\sigma$ increases. It is given in terms of experimental parameters as $\bar I_m/\sigma=\sqrt{2\bar n \eta \kappa T_m}\sin(\vartheta/2)$, where $\vartheta=2\arctan{\chi/\kappa}$.  

  The pulse sequence for determining measurement back action  is shown in Fig. 3A. We first strongly read out the qubit with a 240~ns,  $\bar n=5$ pulse, and record the outcome, which will be used to prepare the qubit in the ground state by post-selection.  Then the qubit is rotated to the $+y$ axis and measured with a variable measurement strength ($T_m=240$ ns), and the outcome $(I_m, Q_m)$ recorded.  The final, tomography, phase measures the $x$, $y$, or $z$ component of the qubit Bloch vector with a strong ($\bar n = 5$, $T_m=240$ ns) measurement pulse.  To compensate for the finite readout strength and qubit temperature, trials with outcomes $|I_m/\sigma| < 1.5$ (corresponding to state purity $<99$~\%) for the first and third measurements are discarded, as well as outcomes for the first measurement with the qubit in $\ket{e}$.  To quantify the measurement back-action for a given measurement outcome $(I_m,Q_m)$, the average final qubit Bloch vector, conditioned by the measurement outcome $(I_m,Q_m)$, $(\langle$X$\rangle_c$, $\langle$Y$\rangle_c$, $\langle$Z$\rangle_c)$, is calculated versus outcome using the results of the tomography phase. These conditional maps of $\langle$X$\rangle_c$, $\langle$Y$\rangle_c$, $\langle$Z$\rangle_c$ were constructed using $201$ by $201$ bins in the plane of scaled measurement outcomes $(I_m/\sigma,Q_m/\sigma)$. 

Results for four measurement strengths increasing by decades from $\bar n = 5\times10^{-3}$ to $5$ are shown in Fig. 3B (see Supplementary Movie S1 of histograms and tomograms for all measurement strengths). The left column shows a two-dimensional histogram of all scaled measurement outcomes recorded during the variable strength readout pulse.  At weak measurement strength, the ground (left) and excited (right) state distributions overlap almost completely. Their separation grows with increasing strength until they are well separated at $\bar n=5$, corresponding to the strong projective measurement shown in Fig. 1A. The rightmost columns show $\langle$X$\rangle_c$, $\langle$Y$\rangle_c$, $\langle$Z$\rangle_c$ versus associated  $(I_m/\sigma,Q_m/\sigma)$ bin. At weak measurement strength ($\bar n \ll 1$), the qubit state is only slightly perturbed, with all measurement outcomes corresponding to Bloch vectors pointing nearly along the $+y$ (initial) axis. However, gradients in $\langle$X$\rangle_c$ along the {$Q_m$-axis} and $\langle$Z$\rangle_c$ along the {$I_m$-axis} are visible, demonstrating the outcome-dependent back-action of the measurement on the qubit state. As the measurement strength increases, so does the back-action, as seen in the increase of the gradients in the $\langle$X$\rangle_c$ and $\langle$Z$\rangle_c$ maps (see Fig. S2).  When the measurement becomes strong, the qubit is projected to +$z$ for positive $I_m$ ({-$z$} for negative $I_m$) while $\langle$X$\rangle$ and $\langle$Y$\rangle$ go unconditionally to zero, as expected.

One of the key predictions of finite-strength measurement theory is that the statistics of the measurement process, in particular the apparent measurement strength in the I-quadrature (which can be determined experimentally from the statistics of the measurement outcomes), are sufficient to infer $z_f$ for any apparent measurement strength or outcome (see Eq.~\eqref{eq:final_bloch_vector_main}).  For weak measurement, where the back-action is symmetric along both $x$ and $z$, the apparent measurement strength determines the amplitude of the $x$ back-action as well (see Supp. Mat. Eq.~(14)). In Fig.~4A, we quantitatively compare this prediction with our experimental result. The scaling coefficients relating measurement outcome to back-action along  $z$, $(\partial \langle $Z$ \rangle_c/\partial I_m)\sigma $, and along $x$, $(\partial \langle $X$ \rangle_c/\partial Q_m)\sigma $, extracted from the tomograms at $I_m=Q_m=0$, are plotted versus the apparent measurement strength extracted from the histograms, $\bar I_m/\sigma$ (see Supp. Mat. sec. 1.4).


Both coefficients, $((\partial \langle $Z$ \rangle_c/\partial I_m)\sigma $ and  $(\partial \langle $X$ \rangle_c/\partial Q_m)\sigma )$, are predicted at $I_m=Q_m=0$ to be equal to $\bar I_m/\sigma$; therefore the data in Fig.~4A should have unity slope. However, finite $T_1$ and $T_2$ acting for a time $\tau$ reduce the state purity and the apparent back-action. To first order, the coefficients are modified to $(\partial \langle $Z$ \rangle_c/\partial I_m)\sigma=(\bar I_m /\sigma) e^{-\tau/{T_1}}$ and $(\partial \langle $X$ \rangle_c/\partial Q_m)\sigma\simeq(\bar I_m/\sigma) e^{-\tau/{T_2}}$ for the $z$ and $x$ back-action,  respectively. In our pulse sequence, $\tau\simeq 380~\text{ns}$, predicting slopes of $0.87\pm0.09$ and $0.58\pm0.06$ for $z$ and $x$, in excellent agreement with the experimentally determined slopes of $0.86\pm0.01$ and $0.55\pm0.01$.  All further theoretical predictions are modified to reflect the effects of $T_1$ and $T_2$, following the description in Suppl. Mater. Eq.~(2). The black curve is the full theoretical dependence of $(\partial \langle $X$ \rangle_c/\partial Q_m)\sigma= \bar I_m/\sigma \cos{\left(\bar Q_m\bar I_m/\sigma^2\left((1-\eta)/\eta\right)\right)}e^{-\bar I_m^2/\sigma^2\left((1-\eta)/\eta\right)}e^{-\tau/T_2}$ using  $
\eta=0.2$, the lowest value of $\eta$ we extract from other measurements (see Supp. Mat. sec. 1.3). We attribute the discrepancy between theory and data at high measurement strength to environmental dephasing effects due to finite $T_2$, and losses before the JPC.  Additionally, we process the tomography results unconditioned by measurement outcome in Fig.~4B. Theory predicts $\langle Y \rangle = e^{ - \bar I_m^2/\eta\sigma^2} \text{cos}\left(\bar I_m \bar Q_m/\eta\sigma^2\right)e^{-\tau/T_2}$.  This expression evaluated  with $\eta=0.2$ is shown as a black curve with the deviation for stronger measurements attributed to dephasing effects due to losses before amplification.  


Similar experiments have studied measurement of the state of a microwave cavity by Rydberg atoms~\cite{Guerlin2007}, and partial nonlinear measurement of phase qubits\cite{Katz2006}.  Also, phase-sensitive parametric amplification has been used to implement weak measurement-based feedback~\cite{Vijay2012}.  In our experiment, the ability to perform both weak and strong high-efficiency, QND, linear measurements within a qubit lifetime, coupled with our high throughput and minimally noisy readout electronics, allow us to acquire $13.5$ billion qubit measurements in approximately $28$ hours, data which can be compared to complete theoretical predictions of the conditional evolution of quantum states under measurement.   They  provide strong evidence that the purity of the state would not decrease in the limit of a perfect measurement, even when the signal is processed by a phase-preserving amplifier.  

Our experiment illustrates an alternate approach to the description of a quantum measurement.
In the case of a qubit, a finite-strength QND measurement can be thought of as
a stochastic operation whose action is unpredictable but known to the experimenters after the fact if they
possess a quantum-noise-limited amplification chain. Any final state is possible, and the type of quantity measured, combined with the measurement strength, determines the probability distribution for different outcomes.  This partial (i.e. finite-strength)
measurement paradigm is not inconsistent with the usual view of projective (i.e. infinite-strength)
measurement. Rather, projective measurement is the limiting case of the broader class of finite strength
measurements.

The finite-strength measurement predictions that we have verified have immediate applicability to proposed schemes for feedback stabilization and error correction of superconducting qubit states. While classical feedback is predicated on the idea that measuring a system does not disturb it, quantum feedback has to make additional corrections to the state of the system to counteract the unavoidable measurement back-action. The measurement back-action that is the subject of this paper thus crucially determines the transformation of the measurement outcome into the optimal  correction signal for feedback. Our ability to experimentally quantify the back-action of an arbitrary-strength measurement thus provides a dress rehearsal for full feedback control of a general quantum system.


\begin{scilastnote}
\item The authors wish to thank Alexander Korotkov, Benjamin Huard, Matthew Reed, Andreas Wallraff and Christopher Eichler for helpful discussions.  Facilities use was supported by the Yale Institute for Nanoscience and Quantum Engineering (YINQE) and the NSF MRSEC DMR 1119826. This research was supported in part by the Office of the Director of National Intelligence (ODNI), Intelligence Advanced Research Projects Activity (IARPA), through the Army Research Office (W911NF-09-1-0369) and in part by the U.S. Army Research Office (W911NF-09-1-0514). All statements of fact, opinion or conclusions contained herein are those of the authors and should not be construed as representing the official views or policies of IARPA, the ODNI, or the U.S. Government. M. M. acknowledges partial support from the Agence National de Recherche under the project EPOQ2, ANR-09-JCJC-0070. M. H. D. acknowledges partial support from the College de France.  S. M. G. acknowledges support from the NSF DMR 1004406.  
\end{scilastnote}

\clearpage


\begin{figure}
\includegraphics{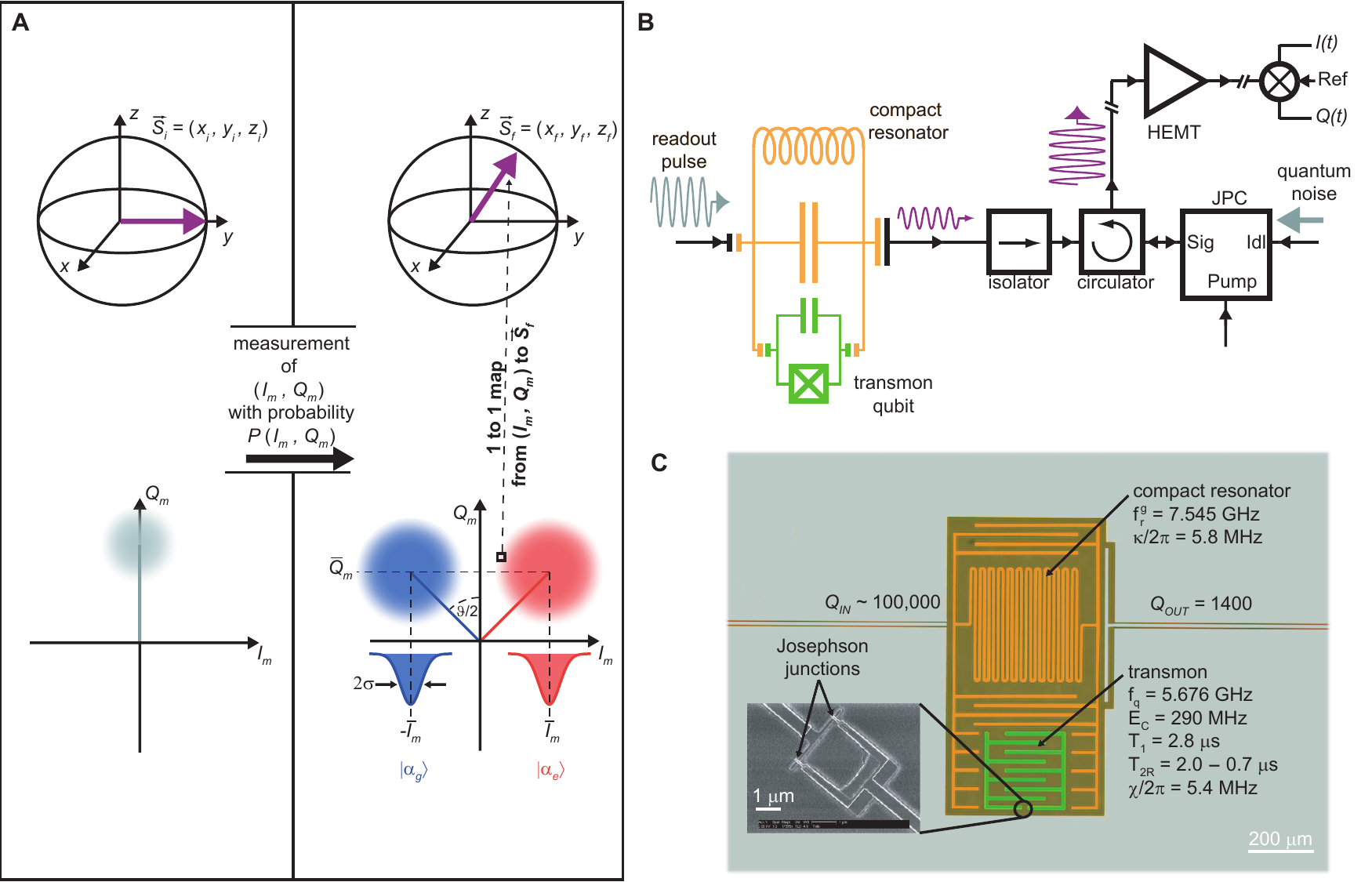}
\end{figure}
\clearpage
\noindent \textbf{Fig. 1. (A)} Bloch sphere representation of the effect on the qubit state of a phase-preserving measurement  in a cQED architecture.  After a measurement with outcome $(I_m,Q_m)$, the qubit will be found in a final state $\vec{S_f}=(x_f, y_f, z_f)$, with $I_m$ encoding information on the projection of the qubit state along $ z$ and corresponding back-action, and $Q_m$ encoding the other component of the back-action, which is parallel to $\hat z \times \vec{S_i} $.  The measurement outcomes are Gaussian distributed, with $\bar I_m^2+\bar Q_m^2=\bar n \kappa T_m$ (see text).
\textbf{(B)} Schematic of experiment mounted to the base plate of a dilution refrigerator. Readout pulses are transmitted through the strongly coupled port of the resonator, via an isolator and circulator, to the signal port (Sig) of a JPC. The idler port (Idl) is terminated in a $50$~$\Omega$ load. The amplified signal output is routed via the circulator and further isolators (not shown) to a High Electron Mobility Transistor (HEMT) amplifier operated at $4$~K, and subsequently demodulated and digitized at room temperature.
\textbf{(C)} False color photograph of the transmon qubit in compact resonator with qubit and resonator parameters.  Inset is a scanning electron micrograph showing the Al/AlO$_x$/Al junction-based SQUID loop at the center of the transmon.

\clearpage
\begin{figure}
\includegraphics{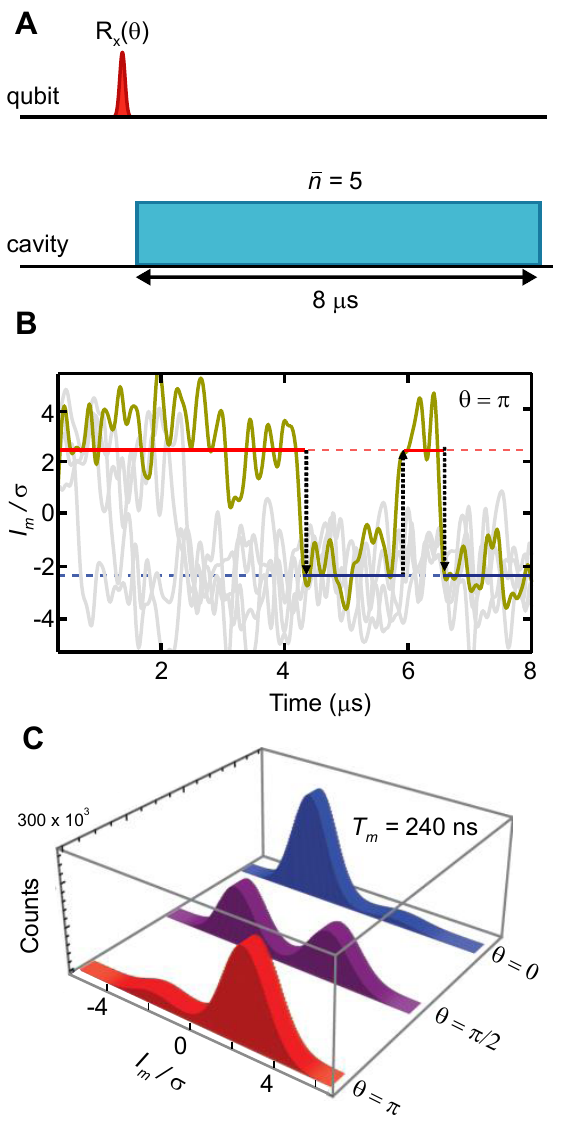}
\end{figure}
\clearpage
\noindent \textbf{Fig.~2.~(A)} Pulse sequence for strong measurement. An initial qubit rotation $R_x(\theta)$ of $\theta$ radians about the x-axis is followed by an $8$~$\mu$s readout pulse with drive power such that $\bar n=5$.  
\textbf{(B)} Individual measurement records.  The data are smoothed with a binomial filter with a $T_m=240$~ns time constant, and scaled by the experimentally determined standard devation ($\sigma$).  Black dotted lines indicate $4\sigma$ deviation events. The qubit is initially measured to be in the excited state, and quantum jumps between excited and ground states are clearly resolved.  The center of the ground and excited state distributions are represented as horizontal dotted lines.    
\textbf{(C)} Histograms of the initial $240$~ns record of the readout pulse along $I_m$ axis, for $\theta=0,\pi/2, \pi$. Finite qubit temperature and $T_1$ decay during readout are visible as population in the undesired qubit state.  

\clearpage
\begin{figure}
\includegraphics{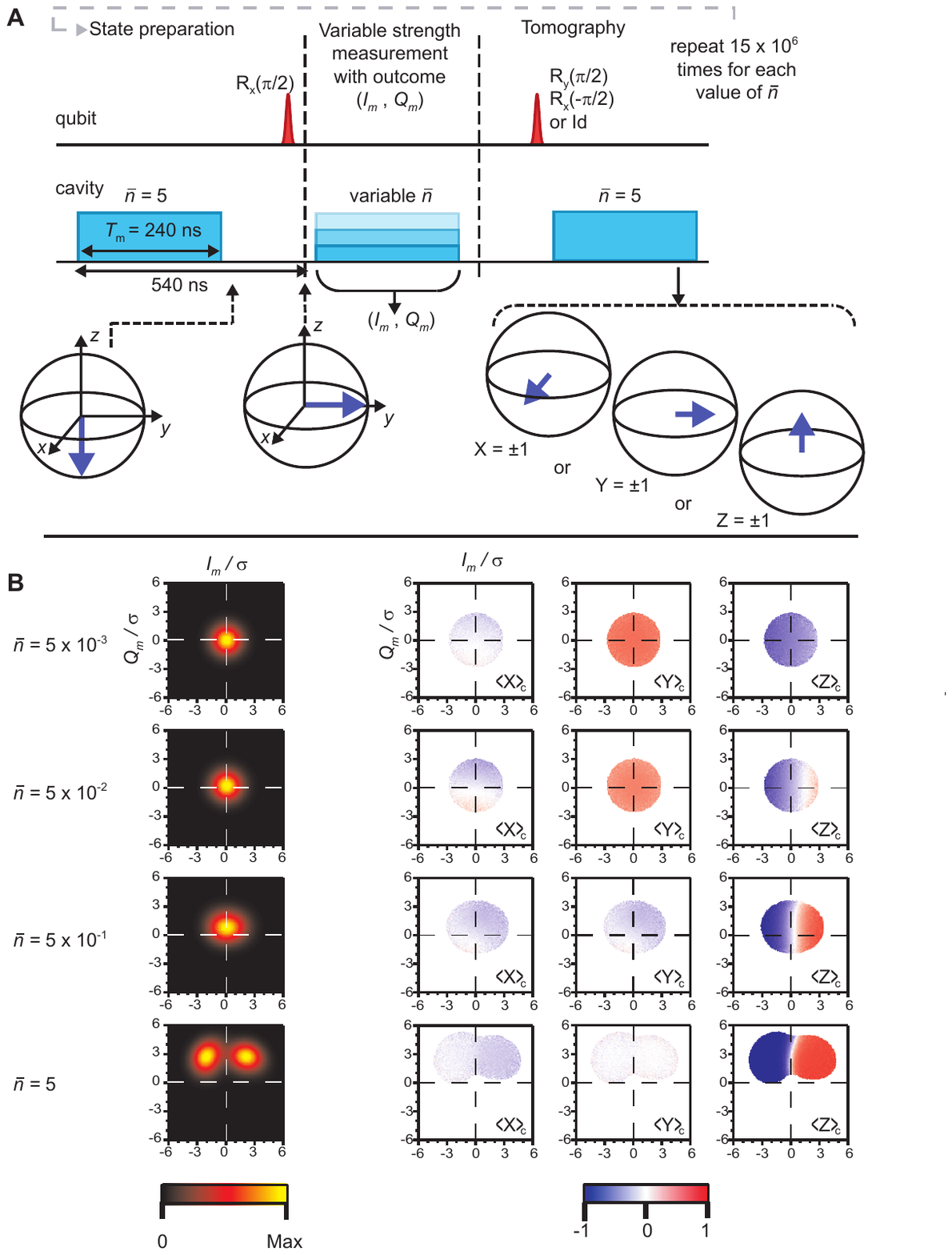}
\end{figure}
\clearpage
\noindent \textbf{Fig.~3.~(A)} Pulse sequence for quantifying measurement back-action.  The measurement strength was varied linearly in amplitude from $\sqrt{\bar n} = 0$ to $\sqrt{5}$. Conditional maps of $\bket{\text{X}}_c$, $\bket{\text{Y}}_c$, $\bket{\text{Z}}_c$ versus measurement outcome $(I_m/\sigma,Q_m/\sigma)$ were constructed using $201$ by $201$ bins.
\textbf{(B)} Results are shown increasing by decades from $\bar n = 5\times10^{-3}$ to $5$. The left column shows a two-dimensional histogram of all scaled measurement outcomes recorded during the variable strength readout pulse. The three rightmost columns are tomograms showing $\langle$X$\rangle_c$, $\langle$Y$\rangle_c$, $\langle$Z$\rangle_c$ versus associated  $(I_m/\sigma,Q_m/\sigma)$ bin. 

\clearpage
\begin{figure}
\includegraphics{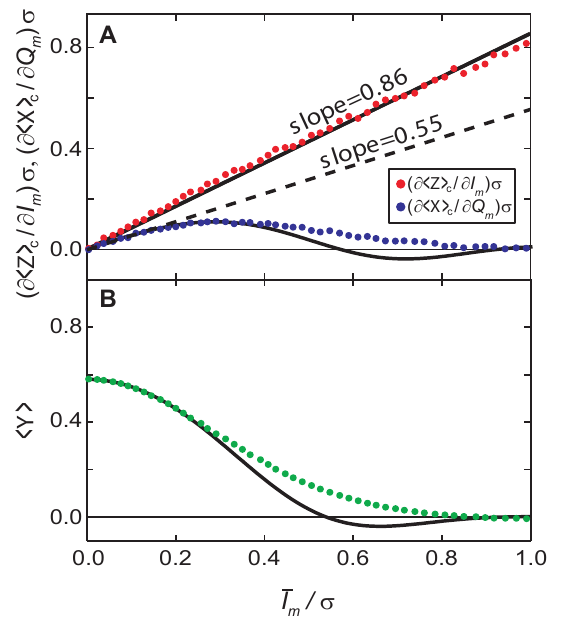}
\end{figure}
\noindent \textbf{Fig.~4.} Correlation between back-action and measurement outcome:
\textbf{(A)} Experimental data for correlated back-action signal along $z$, $(\partial \langle $Z$ \rangle_c/\partial I_m)\sigma $, and along $x$, $(\partial \langle $X$ \rangle_c/\partial Q_m)\sigma $, evaluated at $(I_m,Q_m)=0$, are plotted versus $\bar I_m/\sigma$. For weak measurement strength, the slopes at the origin (represented by solid and dashed line, for $z$ and $x$, respectively) agree with theoretical predictions including first order corrections for $T_1$ and $T_2$.  The solid curve is the full theoretical expression for the $x$ back-action plotted with $\eta=0.2$, $\bar Q_m=1.28 \bar I_m$, and $\text{exp}(-\tau/T_2)=0.58$.
\textbf{(B)} Experimental data for unconditioned $\langle $Y$ \rangle$ versus $\bar I_m/\sigma$.  The data show the expected measurement-induced dephasing when the measurement outcome is not used to condition the perturbed qubit state.  The dephasing rate is proportional to $(\bar I_m/\sigma)^2$, resulting in the apparent Gaussian dependence of $\langle $Y$\rangle $ vs $\bar I_m/\sigma$.  The theoretical expression for $\langle Y \rangle$ vs $\bar I_m/\sigma$ with parameters listed above is shown as a solid curve.

\clearpage

\end{document}